\documentclass[twocolumn]{aastex631}
\usepackage{subfigure}

\usepackage{newtxtext,newtxmath}
\usepackage[T1]{fontenc}
\DeclareRobustCommand{\VAN}[3]{#2}
\let\VANthebibliography\thebibliography
\def\thebibliography{\DeclareRobustCommand{\VAN}[3]{##3}\VANthebibliography}
\usepackage{graphicx}
\usepackage{amsmath}
\usepackage{natbib}
\usepackage{multirow}
\usepackage{subfigure}
\usepackage{hyperref}
\usepackage{textcomp}
\usepackage{rotating}

\newcommand{\Rmnum}[1]{\expandafter@slowromancap\romannumeral #1@}
\makeatother
\begin{document}
\title{Detection of multiple X-ray quasi-periodic oscillations in IGR J19294+1816 with Insight-HXMT}
\author{Wen Yang}
\affiliation{Department of Astronomy, School of Physics and Technology, Wuhan University, Wuhan 430072, China}
\author{Wei Wang}
\altaffiliation{Email address: wangwei2017@whu.edu.cn}
\affiliation{Department of Astronomy, School of Physics and Technology, Wuhan University, Wuhan 430072, China}
\begin{abstract}
We report the timing results with Insight-HXMT observations of X-ray binary IGR J19294+1816 during its 2019 Type \uppercase\expandafter{\romannumeral1} outburst at the decline phase shortly following its peak. We analyze the light curves and power density spectrum (PDS) of the 2019 observations and reveal a peak at approximately $\nu_{NS} \sim 80.2$ mHz, corresponding to X-ray pulsations from the neutron star. In addition, a significant quasi-periodic oscillation (QPO) feature is observed at around $\nu_{QPO} \sim 30.2$ mHz from 10-- 50 keV, with the rms amplitude increasing with energy. Furthermore, we detect two QPOs at the frequency of $\sim 51.1$ mHz and $113.7$ mHz (corresponding to sidebands near $\nu_{NS} \pm \nu_{QPO}$) in 25-50 keV, exhibiting an rms amplitude of around 12\%. Wavelet analysis also shows multiple QPOs at the frequency of $\sim 30$ mHz, $50$ mHz and $ 110$ mHz and these QPO features show transient behaviors, the centroid frequencies of $\sim 30$ mHz remain nearly constant for different luminosities. Our research identifies IGR J19294+1816 as the second strong magnetic-field pulsar with significant sideband signals around the spin frequency. We explore various physical origins that could explain the presence of multiple QPOs.  
\end{abstract}
\keywords{stars: neutron- X-rays: bursts- X-rays: binaries- individual: IGR J19294+1816}
\section{Introduction} \label{sec:intro}
X-ray pulsars in high-mass X-ray binaries (HMXBs) are composed of a highly magnetized rotating neutron star and a high-mass companion star. The accreted matter transfers on to the neutron star unaffected by the highly magnetic field lines until the Alfv\'en radius, where the pressure of the magnetic field balances the ram pressure of the infalling plasma. With magnetic pressure dominating the accreting gas, the flow of infalling gas is channeled along magnetic field lines to the surface of a neutron star, and form “hot spots” near two magnetic poles \citep{shvartsman1971neutron}. A majority of the HMXBs are known to be Be/X-ray binaries (BeXBs) in which young optical companions are spectral type O or B \citep{maraschi1976b}. The X-ray emission in such systems during outbursts is produced when the compact object accretes from a quasi-Keplerian disk around the equator of the rapidly rotating Be star. Such a mechanism explains normal (Type \uppercase\expandafter{\romannumeral1}) outbursts with X-ray luminosity $L_x\,\,\sim~10^{35}-10^{37}\,\,erg\,\,s^{-1}$. Occasionally, they can produce a giant outburst (Type \uppercase\expandafter{\romannumeral1}I), which could occur at any orbital phase and reach a peak luminosity higher than $10^{38}{\,\mathrm{erg~s^{-1}}}$ \citep{reig2011x}.
\par
Timing and spectral properties of radiation generated by accreting compact objects carry information about physical and geometrical properties of these HMXBs. Over the following decades, many observational properties of X-ray pulsars were established, including accreting torque evolution and reversals \citep{malacaria2020ups,mereghetti2015magnetars}, measurements of cyclotron resonance scattering features (CRSFs) in their X-ray spectra \citep{staubert2019cyclotron} and detections of quasi-periodic oscillation (QPO) from power spectra \citep{manikantan2024energy}. Detailed analysis of the emission in different luminosity states allows us to investigate physical processes occurring near the neutron star and at the boundary between the accretion disk and pulsar magnetosphere.
\par
QPOs have been detected in approximately a dozen out of $\sim$100 known accreting X-ray pulsars. These QPOs are primarily concentrated in the low-frequency range from $\sim$ 10 mHz to $\sim$ 1 Hz, which are called mHz QPOs \citep{devasia2011rxte, james2010discovery}. Substantial observational data on mHz QPOs have been accumulated, however, we have no common understandings of the physics origin. The QPOs at 0.2–0.5 Hz in RX J0440.9+4431 occur during the pulse profile's right wing and likely originate from hard X-ray flares \citep{li2024broad,malacaria2024discovery}. Two mHz QPOs in Her X-1, with distinct luminosity dependencies, likely arise from the beat frequency near the $Alfv\acute{e}n$ and corotation radii, and magnetic disk precession, respectively \citep{yang2025observations}. The multiple QPOs in 4U 0115+63 might be caused by instabilities in swirling flows, which are influenced by factors such as viscosity and magnetic fields \citep{ding2021QPOs}. Additionally, QPOs observed in sources such as EXO 2030+375 \citep{angelini1989discovery}, 4U 1901+03 \citep{james2011flares}, V 0332+53 \citep{qu2005discovery}, and SAX J2103.5+4545 \citep{inam2004discovery} would be associated with the rotation of the inner accretion disk. Besides, several key issues remain unresolved, such as the nature of frequency variations, the mechanisms behind the appearance and disappearance of QPOs, and their relationship with less coherent variability. Therefore, further investigation of mHz QPO properties will significantly contribute to advancing theoretical research.
\par
IGR J19294+1816 was first detected during an outburst in 2009 by IBIS/ISGRI instrument aboard the INTEGRAL Gamma-ray Observatory \citep{turler2009integral}. This new BeXB at a distance of $d = 11 \pm 1$ kpc was determined by spectral analysis with infrared photometry in 2018 \citep{rodes2018igr}. Long-term flux variability with an orbital period of about 117.2 days was discovered by the Swift/BAT monitor \citep{2009ATel.2008....1C}. Pulsations at 12.4s were detected from this bright source using Swift observations \citep{rodriguez2009nature}. A QPO feature at $0.032 \pm 0.002$ Hz with an rms fractional amplitude of $\sim 18$ percent was first detected during 2019 type \uppercase\expandafter{\romannumeral1} outburst observed by AstroSat and XMM-Newton, a positive correlation of the QPO rms amplitude with energy is exhibited \citep{raman2021astrosat,manikantan2024energy}. IGR J19294+1816 is also remarkable because it exhibits a QPO as well as a 40 keV CRSF implying a magnetic field strength $4.6\times 10^{12}G$ \citep{tsygankov2019study,raman2021astrosat}.
\par
In this paper, we report the detailed results of the timing analysis of the X-ray pulsar IGR J19294+1816 during the 2019 type \uppercase\expandafter{\romannumeral1} outburst observed with Insight-HXMT. We focus on the QPOs by wavelet analysis and PDS methods. Section \ref{OBSERVATIONS} outlines the observations and data reduction procedures. In Section \ref{DATA ANALYSIS AND RESULTS}, we present the timing results, including the pulse profiles and multiple QPO features identified with the PDS and wavelet methods. The possible physical mechanisms responsible for the QPOs are discussed in Section \ref{DISCUSSION AND CONCLUSIONS}.
\section{OBSERVATIONS}
\label{OBSERVATIONS}
The Hard X-ray Modulation Telescope (Insight-HXMT) is the first Chinese X-ray astronomical satellite launched on 2017 June 15. Insight-HXMT consists of three main instruments: the High Energy X-ray telescope (HE) operating in 20--250 keV and the geometrical areas of the telescopes are 5100 cm$^2$ \citep{liu2020high}, the Medium Energy X-ray telescope (ME) operating in 5--30 keV with a geometrical detection area of 952 cm$^{2}$ \citep{cao2020medium} and the Low Energy X-ray telescope (LE) covering the energy range 1--15 keV with a geometrical detection area of 384 cm$^2$ \citep{chen2020low}. Insight-HXMT was triggered with 6 observations (proposal ID: P0214056) from October 23 to October 30, 2019 during the Type \uppercase\expandafter{\romannumeral1} outburst of IGR J19294+1816 and a total exposure time of $\sim 40$ ks. The Insight-HXMT Data Analysis Software (HXMTDAS) v2.04 is used to analyze data (more details on the analysis were introduced in previous publications, e.g., \citealt{WANG20211}; \citealt{chen2021relation}). To take advantage of the best-screened event file to generate the high-level products including the energy spectra, response file, light curves and background files, we use tasks $he/me/lepical$ to remove spike events caused by electronic systems and $he/me/legtigen$ be utilized to select good time interval (GTI) when the pointing offset angle $< 0.04^\circ$; the pointing direction above earth $> 10^\circ$; the geomagnetic cut-off rigidity $>8$ GeV and the South Atlantic Anomaly (SAA) did not occur within 300 seconds. Tasks helcgen, melcgen, and lelcgen are used to extract X-ray lightcurves with $128^{-1}$ sec time bins. We use the XSPEC v12.12.0 software package \citep{arnaud1996astronomical} included in HEASoft v6.29 for spectral fitting and error estimation. 
\par
\begin{figure}
    \centering
    \includegraphics[width=.48\textwidth]{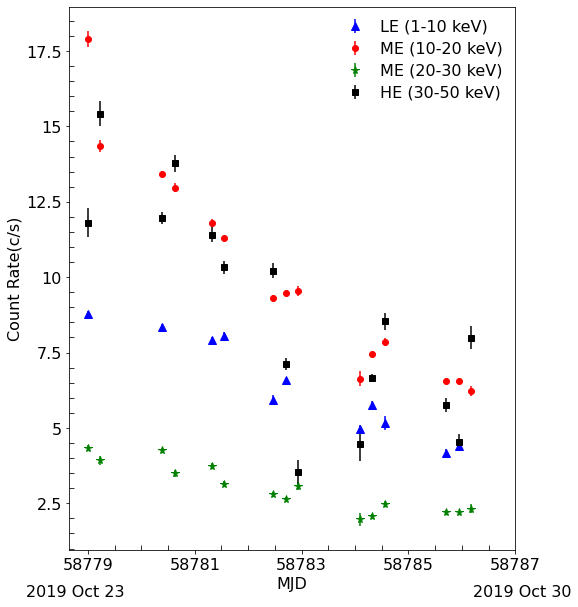}
    \caption{
    Average photon counts per second with a time resolution of 20000s during the Insight-HXMT observations from 2019 October 23 to October 30 for 1-10 keV, 10-20 keV, 20-30 keV, 30-50 keV.}
    \label{fig:1}
\end{figure}

In Figure \ref{fig:1}, the average photon counts per second of IGR J19294+1816 which present the Type \uppercase\expandafter{\romannumeral1} outburst lasting about 7 days monitored by Insight-HXMT are shown, and the pointing observations cover the decrease phase of the outburst. During the observations, the source shows a 2–100 keV luminosity range of $L_x\sim (1.5-3)\times 10^{37}erg\,\,s^{-1}$ at 11 kpc, as estimated by fitting the Insight-HXMT spectra with the model $Tbabs \times cflux\times \left( highecut\times powerlaw \right)$.
\begin{figure}
    \centering
    \includegraphics[width=.42\textwidth,height=0.32\textheight]{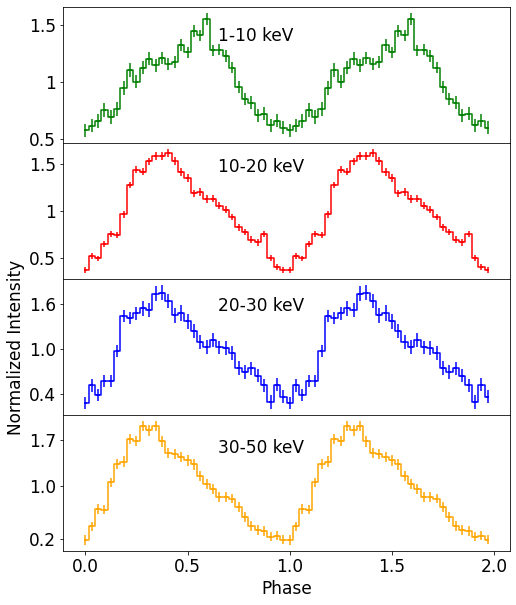}
    \caption{Pulse profiles of IGR J19294+1816 obtained from Insight-HXMT data in different energy bands for ObsID P021405600201.  Zero-phase was chosen arbitrarily to match the minimum in the profile.}
    \label{fig:2}
\end{figure}
\begin{figure*}
    \centering
    \subfigure{
        \includegraphics[width=.48\textwidth,height=20cm, keepaspectratio]{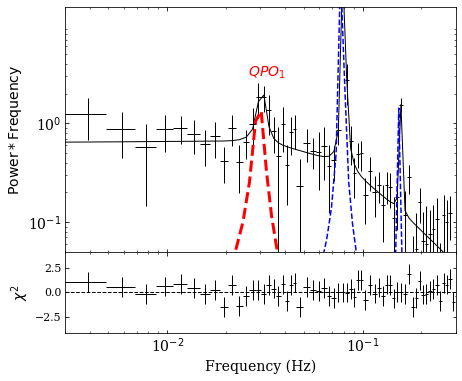}
    }
    \subfigure{
        \includegraphics[width=.48\textwidth,height=20cm, keepaspectratio]{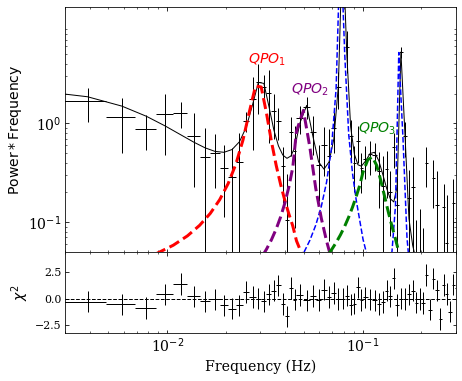}
    }
    \caption{Power density spectrum  of IGR J19294+1816 for the Insight-HXMT observations (ObsID P021405600201) of 10-20 keV (left panel) and 25-50 keV (right panel). The solid black lines correspond to the best-fit model with multi–Lorentzian function. The sharp peaks corresponding to the neutron star's spin period of 12.48 s and its harmonics are indicated by the blue dotted line. The QPO signals, $\nu_{QPO1}\sim 30 mHz$, $\nu_{QPO2}\sim 50 mHz$, $\nu_{QPO3}\sim 110 mHz$, are represented in red, purple and green dotted lines, respectively. }
    \label{fig:3}
\end{figure*}
\section{DATA ANALYSIS AND RESULTS}
\label{DATA ANALYSIS AND RESULTS}
\subsection{Pulse profiles}
The count statistics obtained from Insight-HXMT allowed us to search for the pulsed period with a binning of $128^{-1}$s by using the $efsearch$ method. The uncertainties of the spin period are estimated by folding the light curve with a large number of periods around the approximate period by maximizing $\chi^2$ and determined using the Gaussian function. The spin period is determined at $\sim 12.486(0.004)s$ for all ObsIDs based on the barycenter-corrected light curve. We created the energy-resolved pulse profiles by folding the light curves across different energy bands using the pulse period for each ObsID. As an example of pulse profiles over various energy bands for ObsID P021405600201 are presented in Figure \ref{fig:2}. The entire energy band was resolved into various sub-intervals as: 1-10 keV, 10-20 keV, 20-30 keV, 30-50 keV. The pulse profile displays a broad single peak that spreads across the entire phase range, with an indication of a potential additional component emerging on the right side of the peak. Consistent with previous reports \citep{raman2021astrosat}, a secondary peak, with an intensity approximately 70\% that of the primary peak, located on its left side, is particularly notable below 10 keV. Energy-dependent features in the pulse profile have been observed in many accretion-powered X-ray pulsars, such as 4U 1909+07 \citep{jaisawal2013possible}, 4U 0115+63 \citep{tsygankov20074u}, and EXO 2030+375 \citep{yang2024evidence}. An interpretation suggests that this could be attributed to the observer being expected to detect emission from two distinct regions at low energies but only from a single region at high energies, with low energy X-ray photons being emitted from the upper regions of the accretion column and high-energy photons originating from areas near the neutron star surface \citep{lutovinov2009timing}. Changes in the pulse profile with photon energy can also be affected by local absorption due to asymmetric matter distribution, multiple emission components and gravitational light bending \citep{mushtukov2024accreting}.
\begin{figure*}
    \centering
    \subfigure[]{
        \includegraphics[width=.48\textwidth,height=25cm, keepaspectratio]{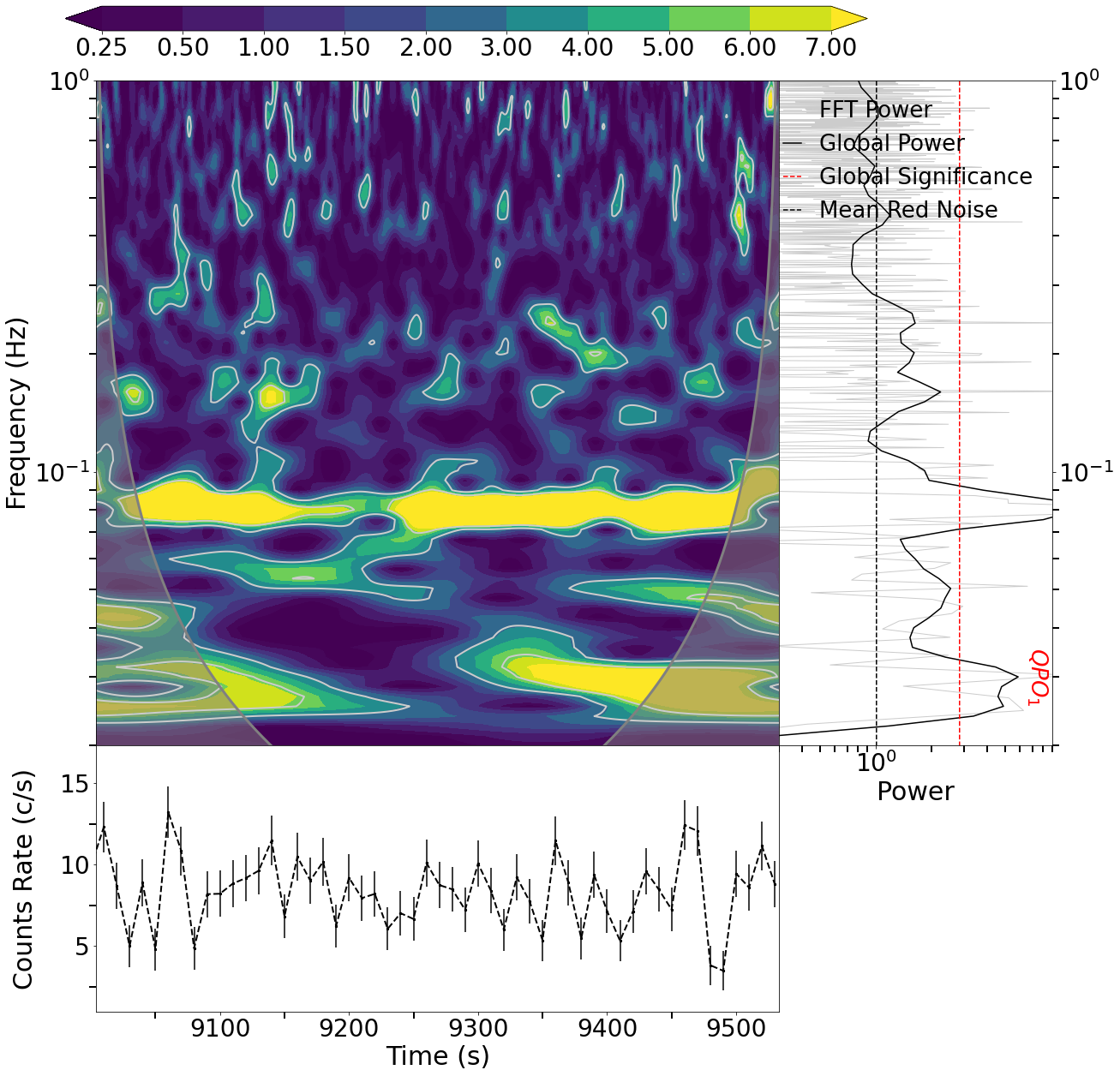}
    }
    \subfigure[]{
        \includegraphics[width=.48\textwidth,height=25cm, keepaspectratio]{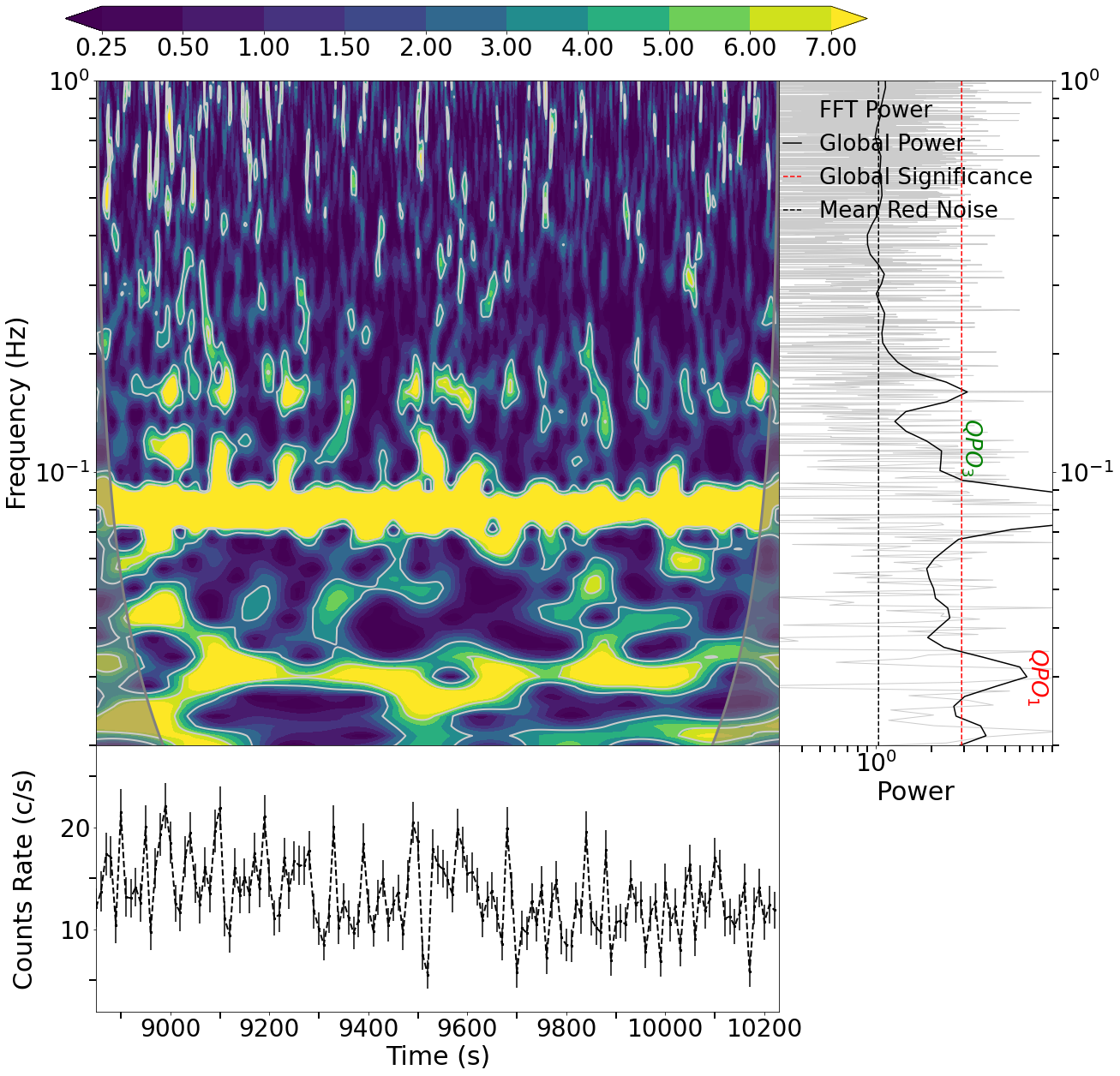}
    }
    \subfigure[]{
        \includegraphics[width=.48\textwidth,height=25cm, keepaspectratio]{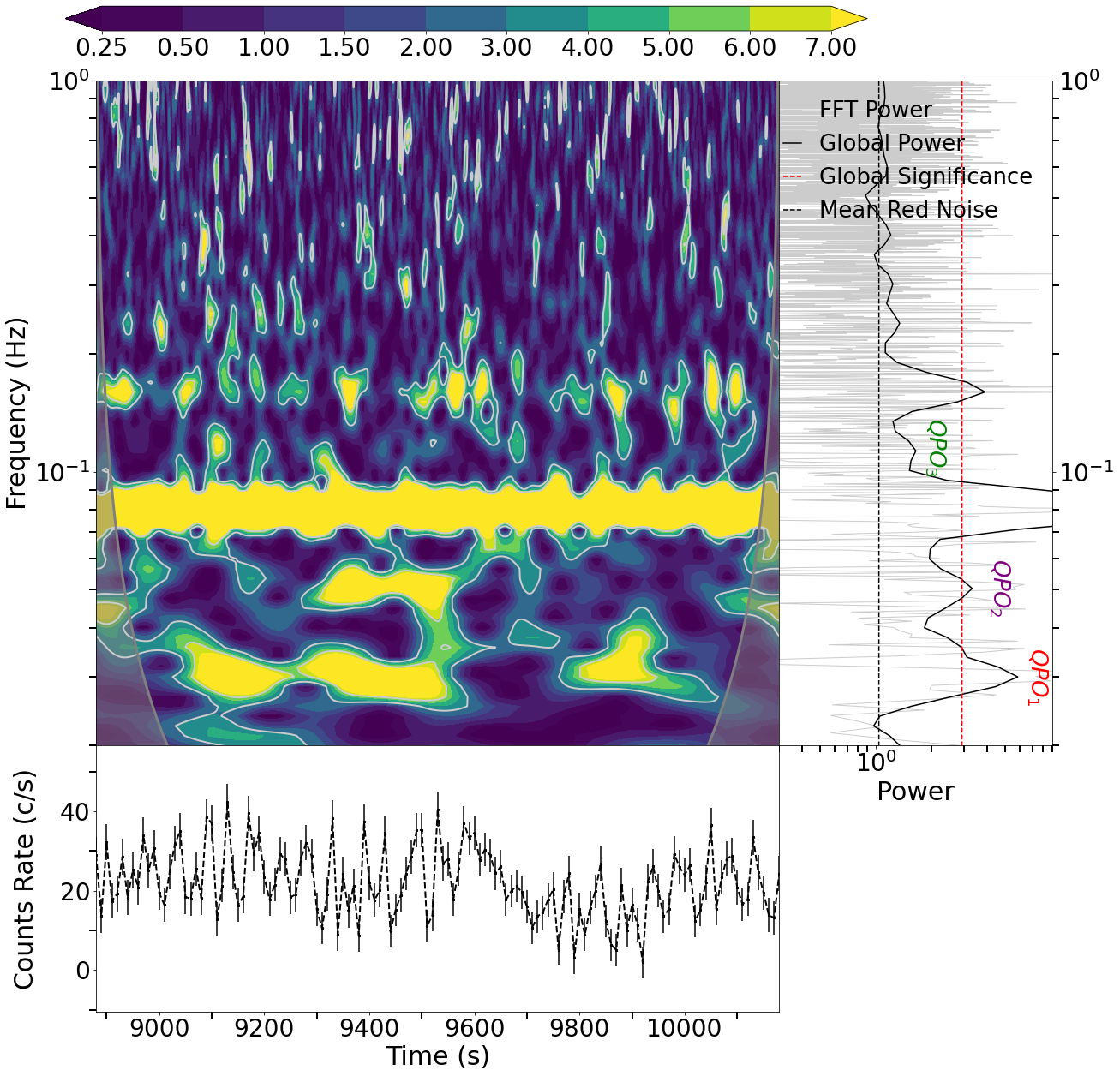}
    }
        \subfigure[]{
        \includegraphics[width=.48\textwidth,height=25cm, keepaspectratio]{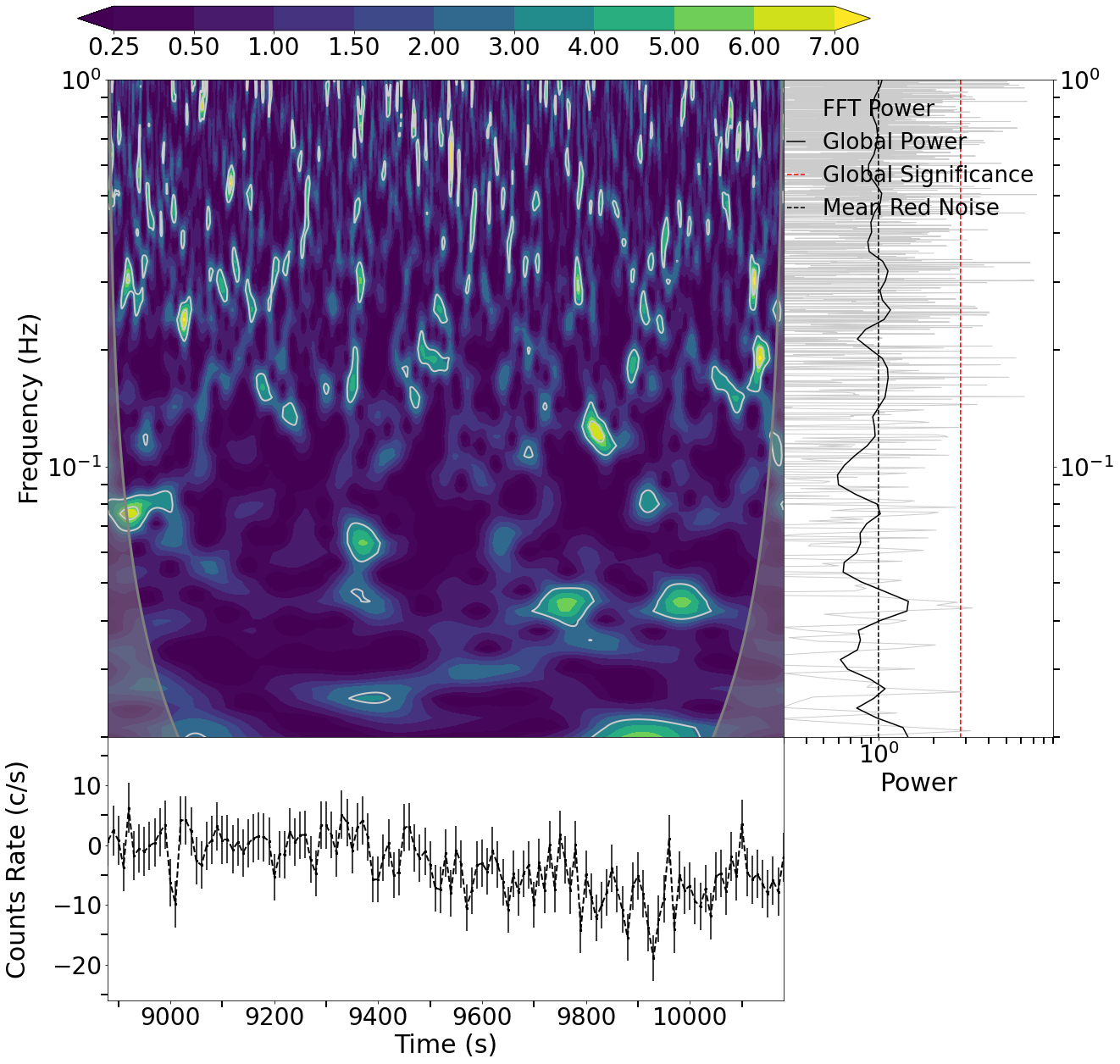}
    }
    \caption{Sub-figures: wavelet power (left) and its global wavelet power spectrum (right), count rates (bottom) with 10 s bins for ObsID P021405600201 in MJD 58781.15 of 1-10 keV (a), 10-20 keV (b), 25-50 keV (c) and 50-100 keV (d). Gray lines in wavelet power refer to the 95\% confidence spectrum and the region where edge effects
    are significant is marked with a gray curved line. A color bar of the contour plot is presented on the top side and the value scale represents the local wavelet power. The solid black line in the global wavelet is the time-averaged power, which is compared to the power spectra of red noise random processes (black broken lines) and red noise at the 95\% significance levels (red broken lines). Three QPO signals of $\nu_{QPO1}\sim 30 mHz$, $\nu_{QPO2}\sim 50 mHz$, $\nu_{QPO3}\sim 110 mHz$ in the wavelet power spectra are labeled in red, purple and green. The $\sim 30$ mHz QPO signals are found below 50 keV. The $\sim 110$ mHz QPO signals were detected in 10-20 keV, both $\sim 50$ and 110 mHz QPO signals were detected in 25-50 keV. Above 50 keV, the background becomes dominant, leading to the disappearance of both the pulsations and the QPO.}
    \label{fig:4}
\end{figure*}
\begin{figure}
    \centering
    \includegraphics[width=.42\textwidth]{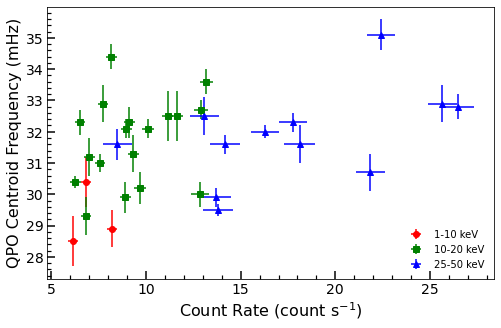}
    \caption{Centroid frequency of $QPO_1$ versus the average count rate corresponding to the GTI data.}
    \label{fig:5}
\end{figure}
\subsection{Power density spectrum}
Inspired by the initial report of the $\sim 30$ mHz QPO in IGR J19294+1816 detected by AstroSat during the 2019 outburst \citep{raman2021astrosat}, we also carefully checked the light curves of IGR J19294+1816 observed by the Insight-HXMT. We first carried out background-subtracted processes on the extracted light curves obtained from each payload and exposure. Powspec from HEASOFT was employed to calculate the PDS for each observation, using a time interval of 512 seconds and a corresponding time resolution of $128^{-1}$ seconds. A final PDS was generated with the average of all the power spectra. The PDSs were normalized to ensure that their integral is equal to the square of the rms fractional variability and rebinned geometrically by a factor of 1.03. To characterize quasi-periodic variability, we carried out the PDS fitting using the XSPEC fitting package. Our PDS model includes several Lorentzian features to characterize a broad-band noise, a spin frequency peak at $\sim 0.08$ Hz, its harmonic at $\sim 0.16$ Hz, and multiple QPO features. The best-fitted parameters obtained from optimal models are presented in Table \ref{tablenew}.
\par
We first investigate the energy dependence of QPO features by creating PDS for different energy bands: 10-20 keV for ME data and 25–50 keV for HE data. Furthermore, the average photon count rate is below 10 counts/s in 1-10 keV for these 6 observations, preventing us from analyzing the PDS in low energy bands. The PDSs for ME and HE observations of ObsID P021405600201 are presented in Figure \ref{fig:3}. The average QPO centroid frequency was determined to be $0.0304\pm0.0006$ Hz and the QPO width has a value of $\sim 0.003$ Hz in 10-20 keV. The QPO centroid frequency is $0.0302\pm0.0001$ Hz with width of $\sim 0.005$ Hz in 25-50 keV. The quality factor $Q=\frac{v}{\bigtriangleup v}$ (where $v$ represents the frequency of the QPO and $\bigtriangleup v$ represents the full width at half maximum; FWHM) is calculated to be $\sim 13$ and $\sim 6$ for ME and HE respectively. Additionally, the white noise-subtracted rms value shows a notable increase, rising from $\sim 8\%$ at 10–20 keV to $\sim 14\%$ at 25-50 keV. A positive correlation of the QPO rms amplitude with energy is consistent with \cite{raman2021astrosat}. We detected $\sim 30$ mHz QPOs beyond 25 keV, whereas such QPOs were not significantly observed above 30 keV previously. For ObsID P021405600401, where the QPO was also detected at 10-20 keV, the average QPO centroid frequency was determined to be $0.0314\pm0.0010$ Hz and the width of QPO $\sim 0.004$ Hz. For other ObsIDs, the lack of clear QPOs may be attributed to the low count rate and the traditional Fourier transform is not sensitive to non-stationary signals.
\begin{table}[htbp]
\centering
\caption{All the QPO parameters obtained from the best-fitting PDS in different energy bands. The ObsID shows the last four digits of P02140560.}
\label{tablenew}
\begin{tabular}{cccc}
\hline
ObsID & Energy (keV) & QPO Parameter & Value \\
\hline
\multirow{12}{*}{0201} 
  & \multirow{3}{*}{10–20} 
      & $\nu_{qpo1}$ (mHz) & 30.4(6) \\
  & & Q factor                       & 12.6(9) \\
  & & rms (\%)                       & 8.4(20) \\ \cline{2-4}
  & \multirow{9}{*}{25–50} 
      & $\nu_{qpo1}$ (mHz) & 30.2(1) \\
  & & Q factor                       & 5.5(4) \\
  & & rms (\%)                       & 15.1(20) \\
  & & $\nu_{qpo2}$ (mHz)  & 51.1(21) \\
  & & Q factor                       & 7.5(4) \\
  & & rms (\%)                       & 12.6(29) \\
  & & $\nu_{qpo3}$ (mHz)  & 113.6(35) \\
  & & Q factor                       & 3.8(1) \\
  & & rms (\%)                       & 14.5(17) \\ 
\hline
\multirow{3}{*}{0401} 
  & \multirow{3}{*}{10–20} 
      & $\nu_{qpo1}$ (mHz) & 31.4(10) \\
  & & Q factor                       & 6.8(3) \\
  & & rms (\%)                       & 15.8(25) \\
\hline
\end{tabular}
\end{table}

At 25–50 keV, the PDS for ObsID P021405600201 shows a new QPO below the main pulsation at $0.0511\pm0.0021$ Hz with width $\sim 0.007$ Hz and rms value of $(12\pm3)\%$. Furthermore, there would be the presence of an additional QPO above 0.05 Hz, which allowed us to add an additional Lorentzian function to the model, with a central frequency at $0.113\pm0.004$ Hz and the width of the Lorentzian function fixed at 0.03 Hz, the $\chi ^2$ values changed from 111 (61 dof) to 85 (59 dof) with a classical F-test probability of $3.8 \times 10^{-4}$ for the model with the addition of a 0.11 Hz QPO. However, the classical F-test could give incorrect results when testing extra components like QPO peak features in power spectra \citep{protassov2002statistics}. To address this problem, we followed the methodology adopted in previous studies \citep{atapin2019ultraluminous} to check the significance. We simulated 10000 PDSs based on the model without the additional Lorentzian component at $\sim 0.11$ Hz, using the XSPEC fakeit command with model parameters fixed at observed best-fitting values. Each simulated PDS was then fitted with two models that add and do not add the Lorentzian component to obtain the simulated F-statistics. By comparing the simulated F-statistics with the observed F-statistics, we derived a corresponding p-value of $2 \times 10^{-4}$. We speculate that this phenomenon is consistent with the Fourier frequency-shifting theorem, which implies that the QPO can modulate the amplitude of the main pulsation, leading to the formation of symmetric sidebands in the power spectrum. This effect has been observed in an X-ray pulsar 4U 1626–67 \citep{kommers1998sidebands,sharma2025sidebands}.
\subsection{Wavelet Analysis}
To test whether the poor approximation for the QPO is due to intrinsic variations of its centroid frequency, we execute the pycwt\footnote{\href{https://github.com/regeirk/pycwt}{https://github.com/regeirk/pycwt}} package to compute the continuous wavelet transform, defined as 
\begin{equation}
W_n\left( s \right)  =\underset{k=0}{\overset{N-1}{\varSigma}}\overset{\land}{x}_k\overset{}{\overset{\land}{\varPsi}^*\left( s\omega_k \right) e^{i\omega_kn\delta t}} ,
\end{equation}
Here, $\overset{\land}{x}_k$ is a discrete Fourier transform of our signal (1-10 keV for LE data, 10-20 keV for ME data, 25-50 keV for HE data with time resolution of $128^{-1} s$), $\overset{\land}{\varPsi}\left( s\omega_k \right)$ is Morle wavelet basis function (see Table 1 of \citet{torrence1998practical}). The scaling parameter s in the wavelet transform is similar to the scaling factor in the Fourier transform, as it represents each frequency component (in our case,  $s_j=s_0\cdot 2^{\left( j-1 \right) \cdot \varDelta j}, j=1,2,3...,\left( \log _2\left( N*dt/s_0 \right) \right) /dj$, N represents the length of the frequency spectrum obtained after performing the Fourier Transform, $s_0$ is the smallest scale of the wavelet, which is twice the value of $128^{-1}$, spacing between discrete scales represented by $\varDelta j$ and default value is 1/12). The shift parameter n can be considered as the time, which is not present within the Fourier transform. A more comprehensive introduction to wavelet analysis and its applications in X-ray light curves can be found in \cite{yang2025observations} and \cite{ghosh2023applying}. Four sections of the wavelet power spectrums are shown in Figure \ref{fig:4}. Three effects can be observed:
\begin{itemize}
    \setlength{\itemsep}{0pt}
    \setlength{\parskip}{0pt}
    \renewcommand{\labelitemi}{}
    \item \noindent —— The QPO appears in the first 100 s and reappears after 200 s in the LE bands. The mHz QPO in the ME bands, with a constant frequency near 30 mHz, persists throughout the entire duration of the GTI. For the HE bands, there is also a QPO near 30 mHz lasting about 600 s in the early stage and another 200 s oscillation in the later stage. 
    \item \noindent —— The light curve shows that the photon count rate varies several times with the onset of the oscillation, as the ME count rate fluctuates between $\sim 10$ and $\sim 20$, and the HE count rate fluctuates between $\sim 10$ and $\sim 30$ (shown in the bottom panel).    
    \item \noindent —— The oscillations as if the $\sim 30$ mHz QPO signal modulate the amplitude of the coherent pulsations were detected at $\sim 50$ mHz and $\sim 110$ mHz in HE bands, and the detections of $\sim 50$ mHz with the 1.2 ks observations reach the expected levels of red noise at the 95\% significance level. The $\sim 110$ mHz QPO signals were detected in 10-20 keV. The 30 mHz QPO lasts longer than the ones at $\sim 50$ mHz and 110 mHz. 
\end{itemize}
We compute the time-averaged wavelet power over the entire GTI to quantify the QPO features globally. The central frequency of the QPO and the FWHM are fitted by the Lorentzian function. The quality factor (Q-factor) and R-factor are estimated accordingly. 
\par
R factor is the power of the global wavelet spectrum relative to the global 95 confidence spectrum:
\begin{equation}
R=\frac{Global\,\,signal\,\,peak}{Global\,\,95\% confidence\,\,spectrum} . 
\end{equation}
To identify significant signals for further analysis, we select QPOs with R-factors exceeding 1.0 and GTIs longer than 500 s. In Table \ref{table1}, we present detailed information for all QPO signals detected in each observation, including the energy band, count rate, exposure time and properties of the $QPO_{1}\sim30$ mHz, $QPO_{2}\sim50$ mHz and $QPO_{3}\sim110$ mHz.

\par
Figure \ref{fig:5} illustrates the relationship between $QPO_{1}$ frequency and average count rate from these GTIs data. We can see that the centroid frequency of the QPO shows variations in the range of 29-35 mHz. The correlation between the two quantities is very weak, with Pearson correlation coefficients of 0.49 and 0.26 in HE and ME, respectively. The average QPO centroid frequency remains consistent in hard X-ray bands, with mean values of $31.7\pm0.5$ mHz for ME and $31.9\pm0.4$ mHz for HE. Below 10 keV, the average QPO centroid frequency is found to be $\sim 29$ mHz. However, the low statistical significance and short GTIs in the LE data only show three data points for detections of mHz QPOs, so that it is difficult to study the QPO centroid frequency variation pattern with different energy bands.
\section{DISCUSSION AND CONCLUSIONS}
\label{DISCUSSION AND CONCLUSIONS}
In this work, we have presented a detailed timing analysis of IGR J19294+1816 during its 2019 Type \uppercase\expandafter{\romannumeral1} outburst by using Insight-HXMT observations. Narrow peaks corresponding to the spin period of the accretion-powered pulsar are clearly seen in PDSs and wavelet spectra. Multiple QPO frequencies at $\sim$ 30 mHz, 50 mHz and 110 mHz were also detected during these observations. The $\sim 30$ mHz QPO feature with $\sim 10$\% rms and quality factor of $\sim 8$ is detected in the power density spectra and wavelet power spectra, and two QPOs at $\sim 50$ mHz and 110 mHz have a quality factor of $\sim 5$ and $\sim 3$ respectively. Here, we present the possible theoretical explanations for the multiple X-ray mHz QPOs in IGR J19294+1816. 
\par
QPOs are traditionally understood to arise from the interaction of matter in the accretion disk with the magnetosphere of a compact object \citep{ghosh1979accretion}. The magnetic stresses are unable to dominate the flow in the accretion disk at radii greater than the $Alfv\acute{e}n$ radius. In neutron star X-ray binaries, the surface magnetic field strength differs between HMXBs and LMXBs. HMXBs generally possess surface magnetic fields on the order of $10^{11}-10^{13}$ G, whereas LMXBs exhibit significantly weaker surface fields, typically around $10^{8}-10^{10}$ G. As a result, in HMXBs, the strong magnetic field remains capable of disrupting the accretion disk even at relatively large radii, thereby giving rise to oscillations at lower frequencies. We can therefore expect to detect mHz QPOs in such systems for both transient and persistent X-ray pulsars \citep{james2010discovery}. In addition, these QPOs can show transient behaviors. \cite{liu2022detection} reported the detection of $\sim 40$ mHz QPOs in Cen X-3 which was not present throughout observations in 2020; a 0.04 Hz QPO in 4U 1626-67 only appeared during the torque reversal to the spin-down state \citep{sharma2025sidebands}; a QPO feature in KS 1947+300 appeared only near the end of the 2001 outburst \citep{james2010discovery}. The QPO features can also reappear and vanish multiple times with almost unchanged luminosity and spectral shape over short timescales (see details in \citealt{ding2021QPOs} and \citealt{yang2025observations}).
\par
Various models have been proposed to explain the characteristics of mHz QPOs in HMXBs, two notable models are the Keplerian Frequency Model \citep{van1987intensity} and the Beat Frequency Model \citep{alpar1985gx5}. In the KFM, QPOs result from the inhomogeneities in the inner accretion disk at the Keplerian frequency. Thus, the Keplerian frequency is $\nu _k=\nu _{QPO}$. However, the KFM is only applicable when the neutron star’s spin is slower than that of the inner accretion disk at the radius where the QPO-generating inhomogeneity is located, since a faster rotating neutron star would generate centrifugal forces that inhibit accretion. The QPO frequency due to BFM is attributed to the modulation in mass accretion rate onto the neutron star’s poles at the beat frequency between the spin frequency and the Keplerian frequency of the inner edge of the accretion disk, according to $\nu _{QPO}=\nu_k-\nu _{spin}$.
\par
For IGR J19294+1816, the spin frequency \(\nu _{\text{spin}} = 80\) mHz is higher than the QPO frequency range (\(\nu _{QPO_1} = 29\)-35 mHz), meaning that the KFM is not applicable in this case.  Then we consider the QPO generation occurring near the Alfv\'{e}n radius \(R_A\) \citep{ghosh1979accretion}, to check whether the BFM is suitable. The value of $R_A$ is estimated with assuming a magnetic field strength of \(4.6\times 10^{12}\) G \citep{raman2021astrosat}, a measured 2-100 keV luminosity ranging from \(L_x=(1.5-3.0)\times 10^{37}\) erg s\(^{-1}\), a neutron star radius of \(10^6\) cm, and a mass of \(1.4 M_{\odot}\), $\Lambda = 1$ for spherical accretion and $\Lambda = 0.1$ for disk accretion \citep{becker2012spectral}.
Under these conditions, the Alfv\'{e}n radius is found to be in the range of $\sim$\(4.8\times 10^7\) cm to \(5.8\times 10^8\) cm. The radius corresponding to the Keplerian frequency at 110 mHz as the inner edge of the accretion disk is $\sim(7.1\text{--}7.3)\times 10^8$ cm. Given the inherent uncertainties and model-dependent assumptions in estimating $R_A$, the discrepancy between these two radii may be not significant. While, frequencies of $QPO_1$ detected within the range of 29-36 mHz (see Figure \ref{fig:5}) do not vary with the source luminosities and the occurrence of $\sim 30$ mHz QPO is transient and lasts longer than the 110 mHz QPO. These properties pose a challenge for the BFM. Therefore, we consider alternative explanations for the generation of the QPOs.
\par
Another possible explanation considering such low-frequency mHz QPOs is the magnetic disk precession model suggested by \cite{shirakawa2002precession}. In this scenario, the inner region of the accretion disk experiences magnetic torques that can induce warping and precession of the disk. Under typical conditions in X-ray pulsars, these  torques can overcome viscous damping, allowing the precessional instability to develop and potentially give rise to mHz QPOs. The precessional frequency of the QPO, as described by Equation (27) in \cite{shirakawa2002precession}, is: 
\begin{equation}
t_{\mathrm{prec}}\simeq 
775.9\left(\frac{L x}{10^{37} \mathrm{erg} \mathrm{~s}^{-1}}\right)^{-0.71}\left(\frac{\alpha}{0.1}\right)^{0.85} \rm s,
\label{equation11}
\end{equation}
where $\alpha$ is the accretion disk viscosity parameter, and $L_{37}$ represents the X-ray luminosity in units of  $10^{37}$ erg s$^{-1}$.  For this calculation, We assume an $\alpha$ =0.023, a value chosen based on the model fits performed by \citet{roy2019laxpc} for the source 4U 0115+63. The obtained value $\sim 10$ mHz is smaller than $\sim 30$ mHz.
\par
The detection of two QPOs at 50 mHz and 110 mHz is not a harmonic of the primary 30 mHz QPO but follows the relationship $\nu _{spin}\pm \nu _{qpo}$ during the decline phase of the 2019 outburst. Similar features were observed in the 4U 1626-67 (\citealt{kommers1998sidebands},\citealt{sharma2025sidebands}). They suggested that sidebands arise due to the Fourier frequency-shifting theorem: for a simple sine wave $\cos \left( 2\pi n_st \right)$, if the amplitude of this sine wave $f\left( t \right)$ changes over time with frequency $n_v$, the Fourier transform will no longer have a single "spike" but will also have additional frequency components around the central frequency $n_{s}\pm n_v$ forming sidebands. The presence of the symmetric sidebands at $\sim 50$ mHz and 110 mHz suggests that the instantaneous amplitude of the coherent pulsations contains a term proportional to the $\sim 30$ mHz QPO signal. The sideband structure provides remarkably constraining information on the $\sim 30$ mHz QPOs. We no longer insist that the QPO frequency corresponds to the Keplerian frequency at the inner edge of the disk and follow the suggestion raised from \cite{kommers1998sidebands}, which explains the presence of QPO signals, symmetric sidebands in a pulsar system. We also propose that a large, coherent structure (referred to as a “blob”) of material orbits the neutron star at a frequency approximately matching the QPO centroid frequency around 30 mHz. During each orbit, a portion of the blob occasionally passes through the line of sight between the neutron star and the observer. This occasionally quasi-periodic reduction in pulsar beam intensity produces transient symmetric sidebands centered around the spin frequency. As the blob moves along its orbit, it absorbs X-rays from the pulsar beam and then scatters them out of the line of sight, giving rise to the direct 30 mHz QPO signal. From our wavelet analysis, sidebands appear more prominently in the HE band, which may be due to the fact that the power of the sidebands typically depends on the amplitude of the modulated signal. The pulse profile shows a higher amplitude in the HE band (see Figure \ref{fig:2}), making the modulation of the pulse beam by the blob more significant. It remains unclear why a single reprocessing structure with limited spatial extent would persist within the accretion disk despite the differential rotation of nearby Keplerian orbits and how this structure modulates the pulsar beam with a quasi-periodic at $\sim 30$ mHz, exhibiting dissimilar behavior for the sidebands over short time scales.
\par
To study the origins of the QPOs ranging from 29-36 mHz in this source, one could examine whether the intensity of its amplitude is related to the accretion rate. If the amplitude intensity is dependent on the accretion rate, the blob would modulate the pulsar beam and scatter them out of the line of sight, serving as the primary driver of this instability. Further theoretical or simulation studies will be required in the future to better understand the origin of this instability.
\par
\section*{Acknowledgements}
We acknowledge the referee for useful comments and suggestions that improved the manuscript. This work is supported by the the NSFC (No. 12133007) and National Key Research and Development Program of China (Grants No. 2021YFA0718503, 2023YFA1607901). 
\begin{table*}[]
\centering
\renewcommand{\arraystretch}{1.5}
\caption{Summary of global parameters of QPOs observed in IGR J19294+1816 with Insight-HXMT during 2019 outburst. Q: quality factor, R: relative to the global 95\% confidence spectrum. The obsID are the last four digits of P02140560.}
\label{table1}
\resizebox{\textwidth}{!}{%
\begin{tabular}{c|ccccccccccccc}
\hline
ObsID & \begin{tabular}[c]{@{}c@{}}Energy\\ (KeV)\end{tabular} & \begin{tabular}[c]{@{}c@{}}Count\\ Rate\\ ($count\,\,s^{-1}$)\end{tabular} & \begin{tabular}[c]{@{}c@{}}Start\\ (MJD)\end{tabular} & \begin{tabular}[c]{@{}c@{}}Exposure\\ (s)\end{tabular} & \begin{tabular}[c]{@{}c@{}}$QPO_{1}$ Centroid\\ Frequency\\ (mHz)\end{tabular} & Q-factor & R-factor & \begin{tabular}[c]{@{}c@{}} $QPO_{2}$ Centroid\\ Frequency\\ (mHz)\end{tabular} & Q-factor & R-factor & \begin{tabular}[c]{@{}c@{}} $QPO_{3}$ Centroid\\ Frequency\\ (mHz)\end{tabular} & Q-factor & R-factor \\ \hline
\multirow{7}{*}{0201} & \multirow{3}{*}{25-50 keV} & 26.50(81) & 58781.08 & 1297 & 32.8(4) & 6.24(18) & 1.19(10) & - & - & - & - & - & - \\
 &  & 21.84(77) & 58781.15 & 1288 & 30.7(6) & 5.29(19) & 1.76(9) & 51.6(3) & 6.07(8) & 1.22(3) & 109.0(32) & 2.55(19) & 0.55(2) \\
 &  & 25.66(76) & 58781.21 & 1261 & 32.9(6) & 4.57(14) & 1.80(8) & 49.3(3) & 5.41(7) & 1.19(3) & - & - & - \\ \cline{2-14} 
 & \multirow{3}{*}{10-20 keV} & 13.19(34) & 58781.08 & 1320 & 33.6(4) & 6.46(13) & 1.27(7) & 60.1(6) & 3.59(9) & 1.19(2) & - & - & - \\
 &  & 12.86(46) & 58781.15 & 1223 & 30.0(4) & 6.38(17) & 2.13(8) & - & - & - & 104.2(61) & 3.12(24) & 0.85(6) \\
 &  & 12.91(38) & 58781.21 & 1439 & 32.7(3) & 4.36(8) & 2.25(4) & 48.0(7) & 4.52(15) & 1.15(5) & - & - & - \\ \cline{2-14} 
 & 1-10 keV & 8.20(28) & 58781.15 & 529 & 28.9(6) & 5.55(19) & 1.96(11) & - & - & - & - & - & - \\ \hline
\multirow{3}{*}{0301} & 25-50 keV & 22.42(76) & 58782.14 & 1379 & 35.1(5) & 6.75(15) & 1.85(9) & 49.5(4) & 4.30(7) & 0.92(2) & - & - & - \\ \cline{2-14} 
 & \multirow{2}{*}{10-20 keV} & 11.63(31) & 58782.14 & 1097 & 32.5(8) & 3.69(17) & 1.65(9) & - & - & - & - & - & - \\
 &  & 11.14(30) & 58782.21 & 1439 & 32.5(8) & 5.32(24) & 1.25(13) & - & - & - & - & - & - \\ \hline
\multirow{4}{*}{0401} & 25-50 keV & 16.30(75) & 58783.20 & 938 & 32.0(2) & 6.53(8) & 1.31(4) & 58.3(8) & 5.66(16) & 1.11(7) & - & - & - \\ \cline{2-14} 
 & \multirow{2}{*}{10-20 keV} & 9.11(29) & 58783.20 & 1200 & 32.3(5) & 6.87(19) & 1.02(11) & - & - & - & - & - & - \\
 &  & 8.95(30) & 58783.27 & 1200 & 32.1(3) & 5.26(9) & 1.99(4) & - & - & - & - & - & - \\ \cline{2-14} 
 & 1-10 keV & 6.12(26) & 58783.17 & 600 & 28.5(8) & 4.75(21) & 1.79(12) & - & - & - & - & - & - \\ \hline
\multirow{3}{*}{0402} & 25-50 keV & 18.12(82) & 58783.31 & 841 & 31.6(6) & 5.01(10) & 1.34(7) & - & - & - & - & - & - \\ \cline{2-14} 
 & 10-20 keV & 9.33(31) & 58783.38 & 1313 & 31.3(6) & 4.06(14) & 2.53(8) & 50.9(11) & 3.69(23) & 0.81(7) & - & - & - \\ \cline{2-14} 
 & 1-10 keV & 6.84(27) & 58783.38 & 779 & 30.4(8) & 7.79(35) & 1.65(17) & - & - & - & - & - & - \\ \hline
\multirow{5}{*}{0403} & \multirow{2}{*}{25-50 keV} & 13.81(79) & 58783.44 & 1220 & 29.5(2) & 7.19(9) & 1.67(4) & - & - & - & - & - & - \\
 &  & 13.68(81) & 58783.47 & 747 & 29.9(3) & 6.95(13) & 1.30(5) & 50.2(3) & 7.49(7) & 1.47(4) & - & - & - \\ \cline{2-14} 
 & \multirow{3}{*}{10-20 keV} & 8.90(30) & 58783.44 & 1200 & 29.9(5) & 2.13(8) & 1.37(3) & - & - & - & - & - & - \\
 &  & 10.10(33) & 58783.47 & 779 & 32.1(3) & 3.86(7) & 1.21(3) & 54.5(5) & 4.36(8) & 1.01(4) & - & - & - \\
 &  & 9.68(30) & 58783.51 & 1020 & 30.2(5) & 4.95(13) & 2.21(7) & - & - & - & - & - & - \\ \hline
\multirow{4}{*}{0501} & \multirow{2}{*}{25-50 keV} & 13.07(76) & 58784.92 & 1231 & 32.5(6) & 5.08(22) & 1.38(8) & - & - & - & - & - & - \\
 &  & 8.48(76) & 58784.99 & 1276 & 31.6(5) & 4.00(22) & 1.48(4) & - & - & - & - & - & - \\ \cline{2-14} 
 & \multirow{2}{*}{10-20 keV} & 6.49(27) & 58784.92 & 1260 & 32.3(4) & 8.05(20) & 1.45(11) & - & - & - & - & - & - \\
 &  & 7.73(28) & 58784.99 & 1290 & 32.9(6) & 3.92(17) & 1.45(6) & - & - & - & - & - & - \\ \hline
\multirow{4}{*}{0502} & 25-50 keV & 17.75(75) & 58785.10 & 1415 & 32.3(3) & 4.53(8) & 2.04(3) & 44.8(5) & 4.67(10) & 1.23(4) & 111.4(54) & 2.86(31) & 0.67(6) \\ \cline{2-14} 
 & \multirow{3}{*}{10-20 keV} & 7.01(28) & 58785.03 & 1320 & 31.2(6) & 2.60(16) & 1.30(4) & - & - & - & - & - & - \\
 &  & 8.16(28) & 58785.10 & 1439 & 34.4(4) & 7.02(14) & 3.12(9) & 45.9(18) & 3.79(23) & 1.16(7) & - & - & - \\
 &  & 7.56(28) & 58785.17 & 1499 & 31.0(3) & 5.17(10) & 2.00(5) & 48.1(9) & 3.41(19) & 1.11(8) & - & - & - \\ \hline
\multirow{2}{*}{0601} & 25-50 keV & 14.17(77) & 58786.45 & 1138 & 31.6(3) & 8.31(13) & 1.50(8) & - & - & - & - & - & - \\ \cline{2-14} 
 & 10-20 keV & 6.26(28) & 58786.45 & 1169 & 30.4(2) & 7.60(10) & 1.31(5) & - & - & - & - & - & - \\ \hline
0602 & 10-20 keV & 6.82(26) & 58786.64 & 899 & 29.3(6) & 5.96(21) & 1.51(12) & 44.8(6) & 4.81(13) & 1.05 & - & - & - \\ \hline
\end{tabular}}
\end{table*}
\bibliography{sample631}{}
\bibliographystyle{aasjournal}
\end{document}